\documentclass[letterpaper,twocolumn,showpacs,superscriptaddress,floatfix]{revtex4}
\usepackage[latin1]{inputenc}
\usepackage{bm}
\usepackage{color, amssymb, graphicx, amsfonts,epstopdf }
\usepackage{multirow,amssymb,amsbsy,amsmath,epstopdf}
\usepackage{graphicx}
\usepackage{verbatim}


\begin{document}

\title{Semihierarchical quantum repeaters based on moderate lifetime quantum memories}

\author{Xiao Liu}
\author{Zong-Quan Zhou$\footnote{email:zq\_zhou@ustc.edu.cn}$}
\author{Yi-Lin Hua}
\author{Chuan-Feng Li$\footnote{email:cfli@ustc.edu.cn}$}
\author{Guang-Can Guo}
\affiliation{CAS Key Laboratory of Quantum Information, University of Science and Technology of China, Hefei, 230026, China}
\affiliation{Synergetic Innovation Center of Quantum Information and Quantum Physics, University of Science and Technology of China, Hefei, 230026, China}
\date{\today}

\begin{abstract}
The construction of large-scale quantum networks relies on the development of practical quantum repeaters. Many approaches have been proposed with the goal of outperforming the direct transmission of photons, but most of them are inefficient or difficult to implement with current technology. Here, we present a protocol that uses a semi-hierarchical structure to improve the entanglement distribution rate while reducing the requirement of memory time to a range of tens of milliseconds. This protocol can be implemented with a fixed distance of elementary links and fixed requirements on quantum memories, which are independent of the total distance. This configuration is especially suitable for scalable applications in large-scale quantum networks.
\end{abstract}

\maketitle

\section{Introduction}

The distribution of entangled photon pairs over long distances plays an essential role in quantum information science. It is the fundamental requirement for the realization of long-distance quantum communication \cite{qkd} and for the construction of optical quantum networks \cite{network}. Entanglement over long distances could also enable large-scale tests of quantum physics \cite{bell}. A simple approach is to create entangled photon pairs locally and send one of the photons to the distant location through an optical fiber. However, the entanglement distribution rate (EDR) decreases exponentially with increasing length of optical links because of inevitable photon loss in the fiber. For example, the rate of photons drops ten orders of magnitude for 500-km transmission through an optical fiber. In classical telecommunications this problem is overcome through the use of amplifiers. Unfortunately, because of the no-cloning theorem \cite{nocloning}, straightforward amplification is not applicable in quantum communication. In this case, quantum repeaters are required, which can improve the scaling of photon loss from exponential to polynomial in principle \cite{first,review}.

A core idea of many quantum repeater schemes is to create short-distance entanglement between optical memories and then extend the distance of entanglement through entanglement swapping \cite{es,first}. The sequence between quantum entanglement creation (EC) and quantum entanglement swapping (ES) is crucial to enhance the EDR. Many previously proposed quantum repeaters protocols can be classified into two categories: hierarchical quantum repeaters and non-hierarchical quantum repeaters \cite{probabilitysource,review}. In a hierarchical quantum repeater, the EC and ES in the elementary links will be performed in a ``repeat until success'' scheme. The entanglement created earlier will be stored until the neighboring link is ready. In contrast, the EC and ES will be performed at the same time for all elementary links in the non-hierarchical quantum repeater. A highly influential proposal by Duan, Lukin, Cirac, and Zoller, which is widely known as the DLCZ protocol \cite{dlcz}, is a typical hierarchical structure. Various improvements have been put forward over the past few years to make this protocol more practical to implement, such as lowering the stability requirements for the channels via two-photon detections in the EC and ES stages \cite{twophoton1,twophoton2,twophoton3,twophoton4,twophoton5,twophoton6,twophoton7}, reducing multiphoton errors through the use of single-photon sources \cite{twophoton4,twophoton7,single1}, and increasing the distribution rate at the elementary level by using temporal or spatial multiplexing \cite{tmultimode,smultimode}. However, there is still an intrinsic problem with its structure that limits its applications. An essential requirement for DLCZ-like protocols is that one must be able to store the created elementary entanglement until entanglement has been established in the neighboring link, the resulting higher-level entanglement again must be stored until the neighboring higher-level link has been established, and so on. Thus, the necessary storage time of quantum memories must be comparable to the total entanglement distribution time, which is highly challenging for practical quantum memories, especially when the distribution distance is rather long.

%%%%% FIGURE 1%%%%%
\begin{figure*}[t]
\begin{center}
\includegraphics [width= 5.5 in]{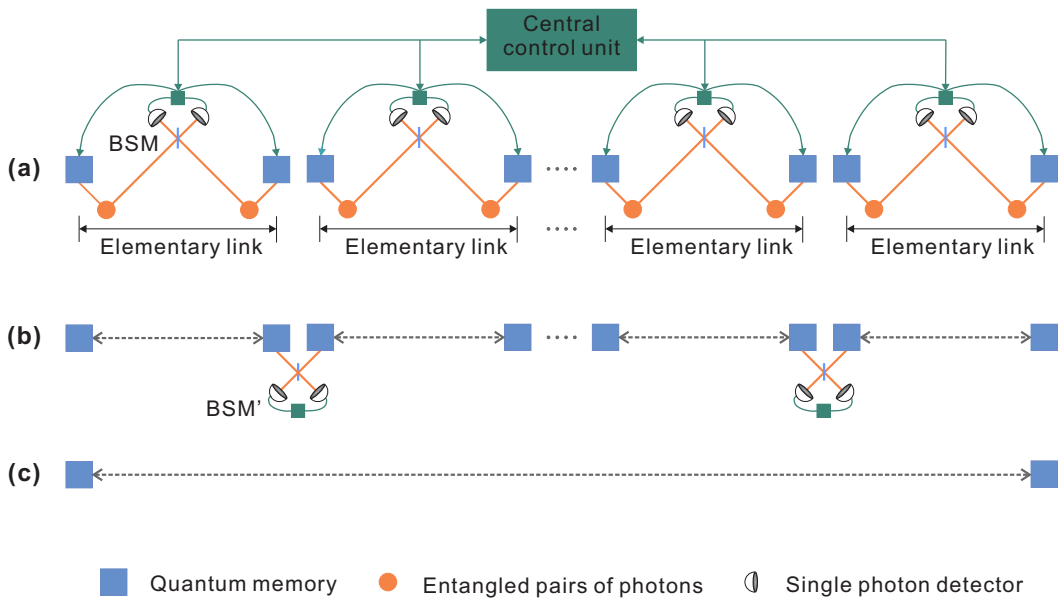}
\end{center}
\caption{(color online) Schematic of the semihierarchical quantum repeater architecture. (a) Both ends of each elementary link contain multimode quantum memories and sources of entangled photon pairs, where each pair is emitted into a different optical mode. One member from each pair is stored in a multimode quantum memory while the other is sent to the center of the link over a quantum channel (orange lines) where they meet photons generated by the sources situated at the other end of the link. Photons occupying the same mode then undergo a mode-resolving Bell-state measurement (BSM) which is composed of a beam splitter (BS) and two single-photon detectors (assuming time-bin qubit encoding and temporal multiplexing is employed). A heralding signal, which originates from a successful BSM, is sent, via classical channel (green lines), to a central control unit located at the midpoint of the total distribution length. The central control unit waits until entanglement (of remote quantum memories)  has been established over each elementary link (depicted by the dotted arrows), and then informs all quantum memories to implement the next step of the protocol. (b) The heralded photons are retrieved from adjacent memories such that they arrive indistinguishably at the BSM' to perform the ES. (c) The entanglement is distributed over the desired distance on the premise that ES operations at all nodes succeed. Otherwise, the whole process must start from scratch.}
\label{scheme}
\end{figure*}
%%%%% FIGURE1 %%%%%

Sinclair \emph{et al}. proposed a non-hierarchical structure repeater \cite{fmultimode} based on an ideal entangled photon source and spectral multiplexing. This protocol does not require a hierarchical connection of the elementary links. With a fixed quantum storage time, both EC and ES processes can proceed simultaneously at each attempt. Thus, the system's clock speed is driven not by the communication time across the elementary link but by the bandwidth of the quantum memory and the number of spectral modes. Despite the appealing performance of this architecture, there are still several significant limitations. This scheme has low tolerance for the number of spectral modes and requires the efficiency of the photon source to be close to unity, which constrains its applications and make it difficult to implement with current technology \cite{probabilitysource}. In addition, there are some non-hierarchical protocols based on encoding and error correction \cite{encoding1,encoding2,encoding3,encoding4}, requiring complicated entanglement resources and gate operations that also make them difficult to achieve.

\section{Semihierarchical quantum repeater protocol}

Here we present a semihierarchical quantum repeater scheme that requires only moderate-lifetime quantum memories and practical photon sources. It combines the advantages of the above two approaches, that is, hierarchical DLCZ-like protocols and non-hierarchical protocols. Specifically, the required memory times are much shorter than those of hierarchical protocols, whereas the requirements for photon sources and the number of modes are lower than those of non-hierarchical protocols.

The diagram of our protocol is presented in Fig. \ref{scheme}. By introducing a central control unit for feedforward control, the ES operations are performed only after the entanglement is successfully established over each of the elementary links. Additionally, owing to the on-demand quantum memory, not only spectral or spatial modes but also temporal modes can be used in this protocol. By introducing more temporal modes, the protocol can be more practically implemented because time-resolved photon emission and detection are well-developed techniques \cite{rou66}. To quantify the performance of our protocol, we calculate the average time required for a successful distribution of an entangled photon pair. Assuming that the total distribution length $L$ is divided into $n$ elementary links, the probability of EC for a single elementary link is given by
\begin{equation}\label{1}
 p = 1 - {\left( {1 - \frac{1}{2}{{\left( {{\eta _D}\rho {{10}^{ - \frac{{\alpha L}}{{20n}}}}} \right)}^{\rm{2}}}} \right)^m},
\end{equation}
where $\rho$ is the emission probability of the entangled photon pair source, with the number of modes denoted by $m$. ${\eta _D}$ is the detection efficiency of single photon detectors.

The whole process is driven by the basic clock interval ${{{L_0}} \mathord{\left/ {\vphantom {{{L_0}} c}} \right. \kern-\nulldelimiterspace} c}$, i.e., the reinitialization time required for an attempt to establish entanglement in elementary links including classical communication, with ${L_0} = {L \mathord{\left/ {\vphantom {L n}} \right. \kern-\nulldelimiterspace} n}$. After an expected time ${T_{ec}}$, when all $n$ elementary links successfully create entanglement through many attempts, the central control unit waits until it receives the signal from every elementary link and then informs all quantum memories to retrieve photons for ES, which takes time ${T_{cc}} = {L \mathord{\left/ {\vphantom {L c}} \right. \kern-\nulldelimiterspace} c}$. A successful entanglement distribution requires ES operations at all nodes to succeed, the probability of which can be denoted by ${P_{es}}$. Thus, the average time we need for an entanglement distribution can be written as
\begin{equation}\label{2}
\begin{aligned}
  {T_{tot}} &= \left( {{T_{ec}} + {T_{cc}}} \right) \times \frac{1}{{{P_{es}}}} \times \frac{1}{{\eta _M^2\eta _D^2}}\\ &= \left( {\frac{{{L_0}}}{c} \times \frac{{f\left( {n,p} \right)}}{p} + \frac{L}{c}} \right) \times \frac{{{2^{n - 1}}}}{{{{\left( {\eta _M^2\eta _D^2} \right)}^n}}}.
\end{aligned}
\end{equation}
with
\begin{equation}\label{3}
  f\left( {n,p} \right) = p\sum\limits_{k = 1}^\infty  {k \times \left\{ {{{\left[ {1 - {{\left( {1 - p} \right)}^k}} \right]}^n} - {{\left[ {1 - {{\left( {1 - p} \right)}^{k - 1}}} \right]}^n}} \right\}},
\end{equation}
where ${\eta _M}$ is the storage efficiency of multimode quantum memories. $\alpha \;{\rm{ = }}\;0.2{{\;{\rm{dB}}} \mathord{\left/ {\vphantom {{\;{\rm{dB}}} {{\rm{km}}}}} \right. \kern-\nulldelimiterspace} {{\rm{km}}}}$ and $c{\kern 1pt} \;{\rm{ = }}\;{\rm{2}} \times {\rm{1}}{{\rm{0}}^5}{{\;{\rm{km}}} \mathord{\left/ {\vphantom {{\;{\rm{km}}} {\rm{s}}}} \right. \kern-\nulldelimiterspace} {\rm{s}}}$ are the attenuation coefficient and speed of telecom wavelength photons transmitted through an optical fiber, respectively. Note that we must wait until all elementary links succeed in establishing entanglement, which takes longer than establishing a single link by a factor of $f\left( {n,p} \right)$. See the Appendix for the detailed derivation.

\section{Simulated performance and comparisons with other protocols}

%%%%% FIGURE 2%%%%%
\begin{figure}
\begin{center}
\includegraphics [width= 1 \columnwidth]{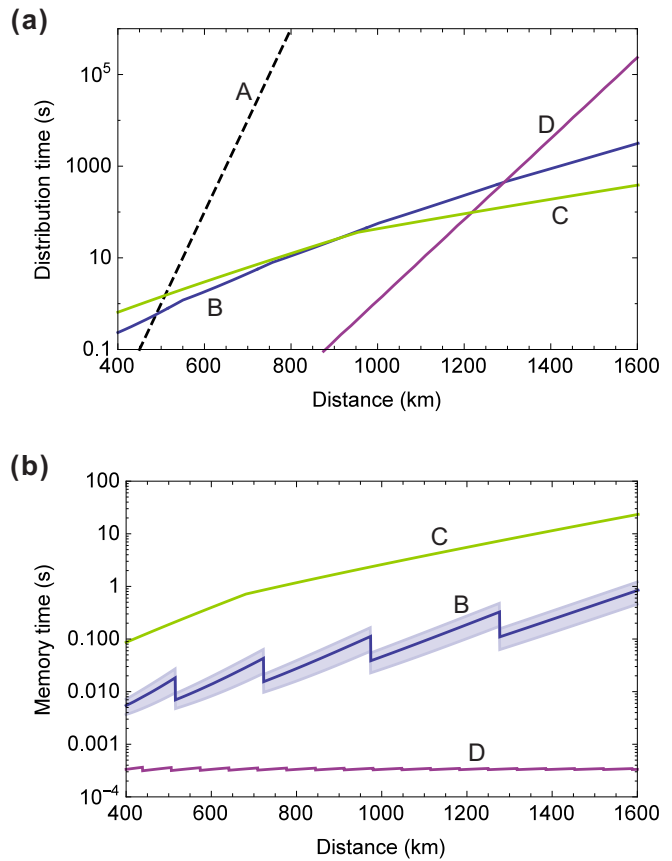}
\end{center}
\caption{(color online) (a) Simulated performance of quantum repeater protocols based on different connection types of the elementary links. The quantity shown is the average time needed to distribute a single entangled pair for the given distance. A: the time required using direct transmission of 10-GHz single photons through fibers. B: the proposed protocol that uses semihierarchical structure. C: the protocol of Ref. \cite{tmultimode}. D: the protocol of Ref. \cite{fmultimode}. (b) The required average memory time of different quantum repeater protocols. The quantity shown is the average memory time needed to distribute a single entangled pair for the given distance. The letters refer to the same protocols as above. For all the curves we have assumed ${\eta _M} = {\eta _D} = 0.9$ and $m = 100$. For curve B and curve D, the emission probability $\rho=0.9$ . The shaded area of curve B represents a one standard deviation uncertainty of required memory time, the detailed derivation of which is contained in the Appendix. All curves are drawn by choosing the optimal link number.}

\label{tot}
\end{figure}
%%%%% FIGURE 2%%%%%

We can calculate the average entanglement distribution time for any distance $L$ and any number of elementary links $n$ by knowing the values of all parameters in the equations above. Assuming that ${\eta _M} = {\eta _D} = \rho  = 0.9$ and $m = 100$, we are able to plot the distribution time as a function of total distance. First, let us consider the shortest distribution time by choosing the optimal value of link number $n$, shown as curve B in Fig. \ref{tot}(a). Direct transmission of a photon, sent over optical fiber, that originates from a source operating at 10 GHz is also included in Fig. \ref{tot}(a) (curve A), as a reference. Our protocol outperforms direct transmission at an approximately 500-km fiber length, and the advantages become increasingly obvious as the distance increases.

We compare the performance of our proposed semihierarchical quantum repeater protocol to other protocols based on different architectures with multiplexing. The average distribution time with respect to distribution distance of these protocols is plotted in Fig. \ref{tot}(a) using similar parameters as assumed in curve B. The protocol of Ref. \cite{tmultimode} (curve C) uses photon-pair sources and multimode memories to implement a temporally multiplexed version of the DLCZ protocol. Due to multiphoton errors, the emission probability of photon pairs is chosen to obtain a final fidelity $F = 0.9$, as in Ref. \cite{tmultimode}. Curve D, which corresponds to the non-hierarchical design protocol proposed in Ref. \cite{fmultimode}, combines two-photon interference and spectral multiplexing. The system's clock speed is driven by the bandwidth of quantum memory and the number of spectral modes. We assume the bandwidth and bandwidth inefficiency to be the same as in Ref. \cite{fmultimode}.

One essential requirement for the quantum repeater protocol is that the quantum memories in each electuary link are required to store the photons until the ES proceeds. In our semihierarchical scheme, we separate the EC process from ES process by using a central control unit. From Eq. \eqref{2}, it is easy to find the average required storage time for memories, written as
\begin{equation}\label{4}
  \left\langle {{t_M}} \right\rangle  = \frac{{{L_0}}}{c} \times \frac{{f(n,p)}}{p} + \frac{L}{c}.
\end{equation}

Using the same parameters assumed above, we can also plot the average memory time $\left\langle {{t_M}} \right\rangle$ as a function of total distance $L$. Fig. \ref{tot}(b) shows $\left\langle {{t_M}} \right\rangle$ as a function of $L$ when link number $n$ is chosen for the optimal distribution time. Owing to our very different structure, the average memory time of the semihierarchical repeater is less than 1 s for a distribution distance of 1600 km, a factor of 27 shorter than that of the hierarchical protocol proposed in Ref. \cite{tmultimode}.

%%%%% FIGURE 3%%%%%
\begin{figure}
\begin{center}
\includegraphics [width= 1 \columnwidth]{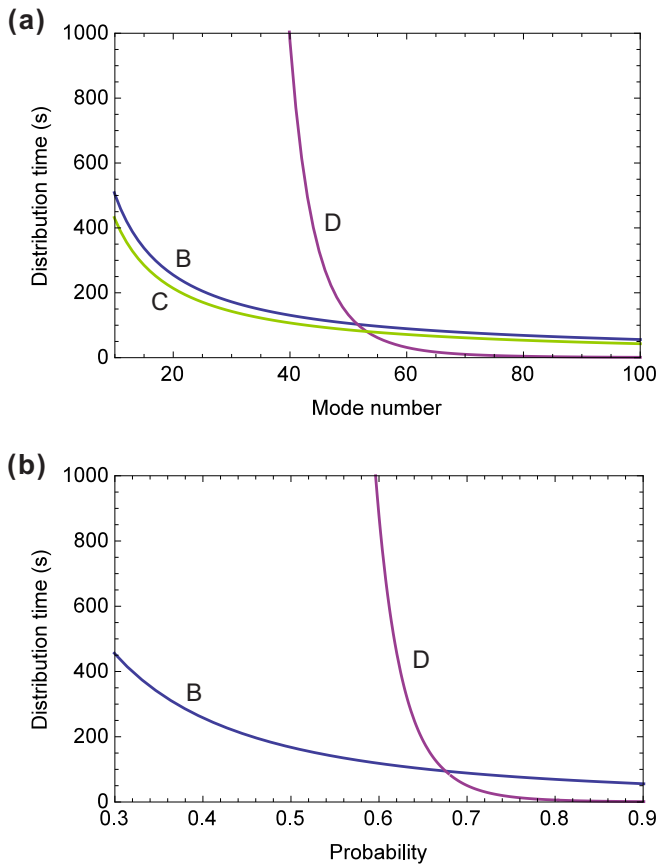}
\end{center}
\caption{(color online) Robustness of different protocols with respect to the reduction of mode number and emission probability. (a) The average time for the distribution of an entangled pair for a distance $L = 1000$ km as a function of the mode number $m$. (b) The average time for the distribution of an entangled pair for a distance $L = 1000$ km as a function of the emission probability $\rho $. Curve C is not included because the pair emission probability of the protocol proposed by Ref. \cite{tmultimode} can not be too high due to multiphoton errors. The letters refer to the same protocols as in Fig. \ref{tot}. All curves are drawn by choosing the best link number to obtain the optimal distribution time.}
\label{reduce}
\end{figure}
%%%%% FIGURE 3%%%%%

Another advantage of our semihierarchical quantum repeater protocol is that the total distribution time has strong robustness for reduced mode number $m$ and emission probability $\rho $ of photon sources. We change either $m$ or $\rho $, while the other parameters remain the same, as assumed before, and then plot the total distribution time as a function of $m$ or $\rho$, respectively. The results are shown in Fig. \ref{reduce}. It is clear that the performance of our proposed architecture remains acceptable even when $m$ drops to 10 or $\rho$ drops to 0.3, which is readily accessible for practical use considering current techniques \cite{temporal3,rou66}.

%%%%% FIGURE 4%%%%%
\begin{figure}
\begin{center}
\includegraphics [width= 1 \columnwidth]{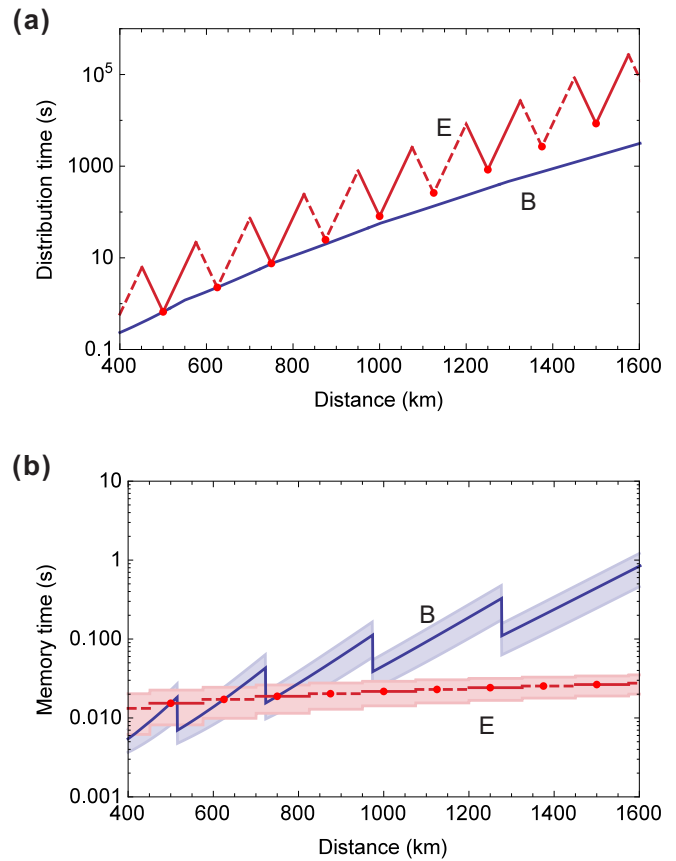}
\end{center}
\caption{(color online) (a) The average time needed to distribute a single entangled pair for a given distance in two scenarios. B: when choosing the best $n$ corresponding to total distance. E: when fixing the length of elementary links to 125 km. The curve E is drawn in two different styles--solid and dashed--and a change in styles means that the optimal link number increments by one. The bolded dots on the curve E represent nodes of the elementary links. Segments with the same style at both sides of one specific bolded dot means that the entanglement is transmitted via fibers from this node. (b) The memory time needed to distribute a single entangled pair for the given distance in two scenarios. B: when choosing the best $n$ corresponding to total distance. E: when fixing the length of elementary links to 125 km. The shaded areas represent the uncertainty of required memory time.}
\label{fixed}
\end{figure}
%%%%% FIGURE 4%%%%%

Note that $f(n,p)$ grows very slowly as link number increases and $p$ is determined by the length of elementary link ${L_0}$ (not only link number); as a result $\left\langle {{t_M}} \right\rangle$ is largely determined by ${L_0}$. We can consider another scenario: setting the length of elementary links to an appropriate constant. Within the realistic range, the smaller ${L_0}$ we choose, the smaller $\left\langle {{t_M}} \right\rangle$ we can obtain. Meanwhile, the total distribution time ${T_{tot}}$ relies on the values of ${L_0}$ as well. Thus, we can obtain a fine tradeoff among ${t_M}$ and ${T_{tot}}$ by setting ${L_0}$ to 125 km to get shorter memory time and reasonable distribution time. The distribution time as a function of total distance is drawn as curve E in Fig. \ref{fixed}(a), with all other parameters as before. The bolded dots on the curve represent $L$ that satisfy $L = n{L_0}$, which are nodes of the elementary links. For $L$ that is not an integer multiple of ${L_0}$, we can distribute the entanglement to one side of its nearest nodes (which side of the nodes is determined by the total distribution time) and transmit the entangled photons to the desired distance through optical fibers. The curve E in Fig. \ref{fixed}(b) shows required memory time as a function of $L$ for ${L_0} = 125$ km. Compared with curve B, the distribution time of which is optimized by choosing the best $n$ corresponding to different $L$ (i.e., ${L_0}$ is not a constant), $\left\langle {{t_M}} \right\rangle$ of curve E is greatly reduced and has little dependence on $L$. For instance, the storage time needed for $L = 1600$ km is approximately 26.5 ms, which is a nearly 32-fold decrease compared with the 0.84 s of curve B. Moreover, the locations of repeater stations (nodes) are fixed for different distribution distances if we fix the length of elementary links, which is an economical way to build a real-word quantum repeater and is suitable for future scalable applications.

\section{Implementation}

We will now briefly discuss the implementation of our proposed scheme. Multiplexed quantum memories with on-demand read-out can be realized based on rare-earth doped crystals since they have excellent coherence properties at cryogenic temperatures, while also providing strong light-matter coupling through high number densities. They also exhibit large static inhomogeneous broadening of the optical transitions, which can be tailored and used as a resource for temporally \cite{temporal1,temporal2,temporal3} and spectrally \cite{fmultimode} multiplexed quantum memories. At the single photon level, on-demand quantum memories have been demonstrated tens of microsecond storage times in praseodymium doped yttrium orthosilicate (YSO) crystal \cite{spinwave1} and millisecond storage times in europium doped YSO crystal \cite{spinwave2} using spin-wave atomic frequency combs (AFCs). The latter experiment also demonstrated temporal multimode storage with five modes; this number was further increased to 50 without sacrificing storage time \cite{50modes}. Except for temporal modes, other forms of modes such as spatial modes \cite{spatial1,spatial2} and spectral modes can also be used together to fulfill multimode multiplexing. For example, 100 modes can be realized by two spectral and five spatial modes in addition to ten temporal modes. It is worth mentioning that europium nuclear spins in YSO have showed quantum lifetimes of up to 6 h at cryogenic temperatures by applying a special magnetic field used to induce a magnetic-field insensitive transition when combined with dynamical decoupling \cite{sixhours}. Moreover, the same idea can also be used in praseodymium doped YSO to increase the storage time up to seconds with strong pulses \cite{memory1s,memory1min}. AFC-based multimode quantum memories currently suffer from low memory efficiency. This may be improved by either putting the memory inside a cavity \cite{highe,highe2} or recalling the photons in the backward direction \cite{temporal1}. The ultimate limit of the these materials is far from being reached. We therefore believe that the memory efficiency and storage time can be further improved in the near future to fit the requirements of our proposed protocol.

Notably, an on-demand single-photon source with 66\% extraction efficiency from a resonantly driven quantum dot in a micropillar \cite{rou66} and single-photon detectors with high efficiency up to 95\% and extremely low intrinsic dark counts have already been demonstrated \cite{detector1,detector2}. The combination of all of these developing technologies makes semihierarchical quantum repeaters seem feasible in the near future.

\section{Conclusions}

The quantum repeater protocol we proposed takes advantage of a semihierarchical structure, which improves the overall entanglement distribution time and reduces the required lifetime of quantum memories to a moderate and accessible range. More intriguingly, the demands for the emission probability of entangled single-photon sources and multimode capacity of quantum memories are also significantly reduced compared with non-hierarchical protocols. We discussed two scenarios in the paper. The first one with optimal length of elementary links has the shortest distribution time and the second one with an appropriate fixed length of elementary links makes the required memory lifetime drop to tens of milliseconds and nearly independent of the total distribution distance. This protocol greatly reduces the technical difficulty for the realization of an efficient quantum repeater and is well suited for scalable applications in a large-scale quantum network.

\section{Appendix}
\subsection{The derivation of entanglement distribution time}

Here we present the detailed analytical derivation leading to the quantitative formulas in the paper. The calculation of the average time for the successful creation of an entangled state across the overall quantum channel can be divided into two parts: The first part is the average time needed before the ES step involving the BSM' measurements, including the expected time of the EC process across all elementary links and the classic communication time needed for feedforward control. The second part is the probability of the ES process and final retrieval.

We will first derive the probability of EC for a single elementary link. The total distance $L$ is split into $n$ elementary links, so the length for each elementary link is ${L_0} = {L \mathord{\left/ {\vphantom {L n}} \right. \kern-\nulldelimiterspace} n}$. All sources are probabilistic single-pair entanglement sources with emission probability $\rho$ and mode number $m$. To deterministically create entanglement between the ends of each elementary link, each photon must travel a distance of ${{{L_0}} \mathord{\left/ {\vphantom {{{L_0}} 2}} \right. \kern-\nulldelimiterspace} 2}$ and be detected by a single-photon detector (SPD) for a successful Bell-state measurement (BSM).

Considering a single elementary link, using only one optical mode, the probability of a successful BSM at the center station is given by
\begin{equation}\label{A1}
  {p_{1m}} = \frac{1}{2} \times {\left( {{\eta _D}\rho {{10}^{ - \frac{{\alpha L}}{{20n}}}}} \right)^{\rm{2}}}, \tag{A1}
\end{equation}
where ${\eta _D}$ is the detection efficiency of SPDs, the pre-factor 1/2 is the maximum successful probability of BSM with linear optics, and ${10^{ - \frac{{\alpha L}}{{20n}}}}$ is the transmission efficiency corresponding to a distance of ${{{L_0}} \mathord{\left/ {\vphantom {{{L_0}} 2}} \right. \kern-\nulldelimiterspace} 2}$ in telecom wavelength optical fibers.

If each source emits $m$ different optical modes at one attempt, the probability that at least one mode results in a successful BSM, which creates the elementary entanglement, is
\begin{equation}\label{A2}
  p = 1 - {\left( {1 - {p_{1m}}} \right)^m} = 1 - {\left( {1 - \frac{1}{2}{{\left( {{\eta _D}\rho {{10}^{ - \frac{{\alpha L}}{{20n}}}}} \right)}^{\rm{2}}}} \right)^m}. \tag{A2}
\end{equation}

For each attempt, one elementary link must wait the time required for the photons to propagate from the sources to the central station and for the information about the result to propagate back to the memories. Thus the basic clock interval of the elementary creation is ${{{L_0}} \mathord{\left/ {\vphantom {{{L_0}} c}} \right. \kern-\nulldelimiterspace} c}$, where $c$ is the speed of telecom wavelength photons in an optical fiber.

One has an exponential distribution of attempt times $k$ for a success probability $p$, written as
\begin{equation}\label{A3}
  P\left( k \right) = {\left( {1 - p} \right)^{k - 1}}p, \tag{A3}
\end{equation}
which gives an expectation value of
\begin{equation}\label{A4}
  \left\langle k \right\rangle  = \sum\limits_{k = 1}^\infty  {kP\left( k \right)}  = \frac{1}{p}. \tag{A4}
\end{equation}

Thus, the expected time of EC for one elementary link is ${{\left( {{{{L_0}} \mathord{\left/ {\vphantom {{{L_0}} c}} \right. \kern-\nulldelimiterspace} c}} \right)} \mathord{\left/ {\vphantom {{\left( {{{{L_0}} \mathord{\left/ {\vphantom {{{L_0}} c}} \right. \kern-\nulldelimiterspace} c}} \right)} p}} \right. \kern-\nulldelimiterspace} p}$. We wait until $n$ elementary links all successfully create their entanglement. Assume that the expected attempt times is increased to ${{f\left( {n,p} \right)} \mathord{\left/ {\vphantom {{f\left( {n,p} \right)} p}} \right. \kern-\nulldelimiterspace} p}$. We will denote the distribution for this combined attempt times by ${P_n}\left( k \right)$ and its expectation value by $\left\langle {\mathop k\limits^ \sim  } \right\rangle$. One has
\begin{equation}\label{A5}
  {P_n}\left( k \right) = \sum\limits_{i = 1}^n {C_n^i} {\left( {P\left( k \right)} \right)^i}{\left[ {\sum\limits_{k = 1}^{k - 1} {P\left( k \right)} } \right]^{n - i}}, \tag{A5}
\end{equation}
where
\begin{equation}\label{A6}
  \sum\limits_{k = 1}^{k - 1} {P\left( k \right)}  = p\sum\limits_{k = 1}^{k - 1} {{{\left( {1 - p} \right)}^{k - 1}}}  = 1 - {\left( {1 - p} \right)^k}. \tag{A6}
\end{equation}

The $i$th term of Eq. \eqref{A5} represents that $i$ of $n$ elementary links create their entanglement successfully at the $k$th attempt, whereas the other $\left( {n - i} \right)$ elementary links succeed before the $k$th attempt. Using the binomial theorem
\begin{equation}\label{A7}
  {\left( {a + b} \right)^n} = \sum\limits_{i = 0}^\infty  {C_n^i} {a^i}{b^{n - i}}, \tag{A7}
\end{equation}
one finds
\begin{equation}\label{A8}
\begin{aligned}
  {P_n}\left( k \right) &= {\left\{ {\left. {P\left( k \right) + \left[ {\sum\limits_{k = 1}^{k - 1} {P\left( k \right)} } \right]} \right\}} \right.^n} - {\left[ {\sum\limits_{k = 1}^{k - 1} {P\left( k \right)} } \right]^n}\\ &= {\left[ {1 - {{\left( {1 - p} \right)}^k}} \right]^n} - {\left[ {1 - {{\left( {1 - p} \right)}^{k - 1}}} \right]^n}.
\end{aligned} \tag{A8}
\end{equation}

The corresponding expectation value becomes
\begin{equation}\label{A9}
  \frac{{f\left( {n,p} \right)}}{p} = \left\langle {\mathop k\limits^ \sim  } \right\rangle  = \sum\limits_{k = 1}^\infty  k {P_n}\left( k \right), \tag{A9}
\end{equation}
giving
\begin{equation}\label{A10}
  f\left( {n,p} \right) = p\sum\limits_{k = 1}^\infty  {k \times \left\{ {{{\left[ {1 - {{\left( {1 - p} \right)}^k}} \right]}^n} - {{\left[ {1 - {{\left( {1 - p} \right)}^{k - 1}}} \right]}^n}} \right\}}. \tag{A10}
\end{equation}

Thus, the average time to create entanglement across all elementary links is given by
\begin{equation}\label{A11}
  {T_{ec}} = \frac{{{L_0}}}{c} \times \frac{{f\left( {n,p} \right)}}{p}. \tag{A11}
\end{equation}

After each elementary link successfully creates entanglement, a signal is sent to a feedforward control module located at the center of the total distance through a classical channel. The central control unit waits until entanglements are successfully established in all elementary links and then informs all quantum memories to implement the next step of the protocol, which takes time ${T_{cc}} = {L \mathord{\left/ {\vphantom {L c}} \right. \kern-\nulldelimiterspace} c}$.

Once all procedures mentioned above are complete, ES between neighboring links is attempted. This process includes recalling photons from the memories with an overall efficiency of ${\eta _M}$ and a BSM'. If there are $n$ elementary links, there are $\left( {n - 1} \right)$ such ES operations, giving
\begin{equation}\label{A12}
  {P_{es}} = {\left( {\frac{1}{2}\eta _M^2\eta _D^2} \right)^{n - 1}}. \tag{A12}
\end{equation}

The entanglement is distributed over the desired distance on the premise that ES operations at all nodes succeed. Otherwise, the whole process must start from scratch. Finally, the entanglement is useful only if it can be retrieved from the memories at the ends of the quantum channel and detected, so the probability $\eta _M^2\eta _D^2$ must be considered.

Therefore, the average time we need for an entanglement distribution can be given as
\begin{equation}\label{A13}
\begin{aligned}
  {T_{tot}} &= \left( {{T_{ec}} + {T_{cc}}} \right) \times \frac{1}{{{P_{es}}}} \times \frac{1}{{\eta _M^2\eta _D^2}}\\ &= \left( {\frac{{{L_0}}}{c} \times \frac{{f\left( {n,p} \right)}}{p} + \frac{L}{c}} \right) \times \frac{{{2^{n - 1}}}}{{{{\left( {\eta _M^2\eta _D^2} \right)}^n}}}.
\end{aligned} \tag{A13}
\end{equation}

\subsection{The standard deviation of memory time}
From the calculations above, it is easy to find the average storage time required for memories, written as
\begin{equation}\label{A14}
  \left\langle {{t_M}} \right\rangle  = \frac{{{L_0}}}{c} \times \frac{{f(n,p)}}{p} + \frac{L}{c}. \tag{A14}
\end{equation}

If we want to fully access the requirement for the memories of the proposed repeater protocol, it is also necessary to calculate the standard deviation of memory storage time. Consider the distribution of one specific entangled photon pairs. Assume the attempt times of this clock cycle is $k$. We can write the actual storage time of the whole distribution process as
\begin{equation}\label{A15}
  {t_M}\left( k \right) = \frac{{{L_0}}}{c} \times k + \frac{L}{c}. \tag{A15}
\end{equation}

The probability distribution of attempt times is ${P_n}\left( k \right)$ as calculated above. Therefore, the variance of memory time is
\begin{equation}\label{A16}
  D\left( {{t_M}} \right) = \sum\limits_{k = 1}^\infty  {\left[ {{{\left( {{t_M}\left( k \right) - \left\langle {{t_M}} \right\rangle } \right)}^2} \times {P_n}\left( k \right)} \right]}, \tag{A16}
\end{equation}
which gives the standard deviation
\begin{equation}\label{A17}
  \sigma \left( {{t_M}} \right) = \sqrt {\sum\limits_{k = 1}^\infty  {\left\{ {{{\left[ {\frac{{{L_0}}}{c}\left( {k - \frac{{f\left( {n,p} \right)}}{p}} \right)} \right]}^2} \times {P_n}\left( k \right)} \right\}} }. \tag{A17}
\end{equation}

{\bf  Acknowledgments}
This work was supported by the Strategic Priority Research Program (B) of the Chinese Academy of Sciences (Grant No. XDB01030300), National Natural Science Foundation of China (Grants No. 11504362, No. 11274289, No. 11325419, No. 61327901, No.11654002), the Fundamental Research Funds for the Central Universities (Grants No. WK2470000023 and No. WK2470000024), and Key Research Program of Frontier Sciences of the Chinese Academy of Sciences (Grant No. QYZDY-SSW-SLH003).

\end{document}